\documentstyle[12pt,A4]{article}
\begin{document}
\renewcommand{\labelenumi}{[\theenumi]}
\begin{center}
{\bf TOTAL POLARISATION CONVERSION IN TWO-DIMENSIONAL ELECTRON
SYSTEM  UNDER CYCLOTRON POLARITON
RESONANCE CONDITIONS}\\

V. V. Popov$^{*)}$, T. V. Teperik\\
{\sl Saratov Division of the Institute of
Radio Engineering and Electronics,\\
Russian Academy of Sciences,
410019 Saratov, Russia\\
$*)$ e-mail:popov@ire.san.ru, Fax:7(8452)519104}
\end{center}

{\footnotesize
The polarisation conversion of a linear polarized
electromagnetic wave incident onto a two-dimensional ($2D$)
electron system at an angle $\theta$ is theoretically studied.
We consider the $2D$ system located at the interface between two
dielectric media with different dielectric constants.
An external dc magnetic
field is assumed to be directed along the normal to the $2D$
electron layer. In such a configuration the cyclotron-polaritons
(CPs)
in $2D$ electron system can be excited with  the frequencies in
the vicinity of the cyclotron frequency. Under the CPs
excitation the resonance polarisation
conversion of electromagnetic wave greatly increases
in the system. In the absence of the electron
scattering in $2D$ system, the polarisation conversion reaches
100\% at a certain value of the angle of incidence
$\theta>\theta_R$, where $\theta_R$ is the total reflection angle.
Extremely high polarisation conversion takes
place in a quite wide range of variation of the angle of
incidence.  High polarisation conversion efficiency (above 80\%)
remains when the actual electron scattering in the $2D$ system on
GsAs is taken into account. The considered phenomena may be
taken up in polarisation spectroscopy of $2D$ electron systems.}

\vspace{1cm}

It is well known that optical-activity effects, which are optical rotation and
polarisation conversion, take place in the systems which possess no mirror
planes of symmetry. Just such a system is two-dimensional ($2D$)
electron plasma layer in an external magnetic field. The
conversion of the polarisation of an electromagnetic wave (EW) shined
onto magnetoactive $2D$ electron system is usually small [1-3] because in
this case the region of interaction of EW with the polarisation active
medium is small. However, the effect may increase substantially
(resonantly) when the external EW excites eigen-oscillations in the $2D$
system.

The resonant Faraday effect arising in a system of $2D$ electron disks as
result of the excitation of edge magnetoplasmons was observed in [4].
The results obtained in [4] show that the effect becomes several times
stronger under magnetoplasma resonance conditions. At the same time,
the power of the polarisation-converted wave in the experiments [4]
remains comparatively low (less than 10\% of the power of the incident
wave). The smallness of the resonance effect in this case is most likely
due to the mismatch of the distributions of the field of the external
(uniform in the plane of the $2D$ system) EW and the
field of nonuniform edge magnetoplasmons, as a result of which the
excitation efficiency of the latter decreases substantially.

The resonant polarisation conversion of an electromagnetic wave
which is assosiated with
the excitation of uniform transverse plasma oscillations of
electrons in a thin semiconductor film has been studied in [5]. The
magnitude of the effect is proportional to a small parameter $d/\lambda$,
where $d$ is the film thickness and $\lambda$ is the EW wavelength.
Naturally, this lowers the efficiency of resonant
polarisation conversion in thin layers.

The limitations due to a small thickness of the electron layer, in
principle, do not arise if the external EW excites
in-plane eigen-oscillations in the layer.
There are two types of in-plane
collective excitations in a homogeneous $2D$ electron system.
They are
well known surface magnetoplasmon-polaritons (MPPs) and
cyclotron-polaritons (CPs) [6].
In [7] it is shown that when the MPPs are excited in a $2D$ electron
system by an external EW in an attenuated total
reflection (ATR) geometry, almost complete conversion of the incident
$p(s)$-polarised wave into a wave with $s(p)$-polarisation is
possible. However, according to [7], total polarisation conversion
occurs in the limit of weak coupling of the external EW
with $2D$ electron system. In this case, two resonance conditions,
corresponding to excitation of the both MPPs and cyclotron oscillations
in $2D$ electron layer must be fulfilled at the same time.
It is clear that these conditions lead to very stringent
requirements on the parameters of a possible expirement
on the observation of total polarisation conversion.
Moreover, dissipation of the energy of the EW due to electron
scattering in a real $2D$ system can lead to almost complete
distraction of a weak coupling of the external EW
with the oscillations of a $2D$ electron plasma in the
ATR geometry.

In the present work the phenomenon of the resonant polarisation
conversion of EW which excites the CPs in a $2D$ electron system
is studied theoretically. Since the CPs are radiative oscillations,
they can be excited by the incidence of the external EW
directly onto the surface of a $2D$ system without using
any additional devices (like the ATR structures).
In this situation the  CPs in a $2D$ system
are strongly coupled to the incident EW, greatly expanding
the possibilities of observing the phenomena considered below.

In the structure considered here, the $2D$ electron layer
is located
in the $x-y$ plane in the interface of two media with
dielectric conctance $\varepsilon_1$ and $\varepsilon_2$.
Static magnetic field $\vec B_0$ is directed along the $z$
axis.
One can solve the whole magneto-optical problem in the structure shown
in Fig.1 starting from the Maxwell equations with boundary conditions
at the interface of the media which take into account the response of the
magnetoactive $2D$ electron plasma. We describe the $2D$ electron
plasma in a simpliest local Drude model for conductivity tensor
\begin{displaymath}
\hat \sigma=
\left(
\begin{array}{r}
\sigma_{\perp}\quad\sigma_{\times}\\
-\sigma_{\times}\quad\sigma_{\perp}\\
\end{array}
\right)
\end{displaymath}
with the components
\begin{equation}
\sigma_{\perp}=\sigma_0 \frac{1-i\omega\tau}
{(\omega_c\tau)^2+(1-i\omega\tau)^2},\qquad
\sigma_{\times}=-\sigma_0
\frac{\omega_c\tau}
{(\omega_c\tau)^2+(1-i\omega\tau)^2},
\end{equation}
where $\omega$ is the angular frequency of the EW,
$\omega_c=|e|B_0/cm^*$ is the cyclotron frequency,
$\sigma_0=e^2N_s\tau/m^*$ is the conductivity of the $2D$ electron
system in the absence of an external magnetic field,
$e$ and $m^*$ are, respectively,  the charge and effective mass of
electron, $N_s$ is the surface electron
density, $\tau$ is the phenomenological relaxation time of the electron
momentum in $2D$ system.

Let us assume that the EW incident onto $2D$ system
from  medium {\sl 1} (see Fig.1)
at an angle $\theta$ to the $z$ axis
has a linear $p$-polarisation so that
the electric vector of the wave resides in the plane of incidence
($x-z$ plane).
Then we
can introduce the magneto-optical complex coefficients for electric
fields:
\begin{equation}
r_{pp}=\frac{E_p^{(r)}}{E_p^{(i)}},\qquad
r_{sp}=\frac{E_s^{(r)}}{E_p^{(i)}},\qquad
t_{pp}=\frac{E_p^{(t)}}{E_p^{(i)}},\qquad
t_{sp}=\frac{E_s^{(t)}}{E_p^{(i)}},
\end{equation}
where superscripts $i$, $r$, and $t$ refer to the incident, reflected, and
transmitted waves, respectively, and the subscripts $p$ and $s$
correspond to waves with $p$ and $s$ polarisation (in the latter case the
electric field vector of the EW is perpendicular to the plane of incidence).
We can also introduce the power conversion coefficients as ratios of the
components of the energy flux vectors normal to the plane of the $2D$
system:
\begin{equation}
\begin{array}{l}
R_{pp}=\frac{\displaystyle P_{pz}^{(r)}}
{\displaystyle P_{pz}^{(i)}},\qquad
R_{sp}=\frac{\displaystyle P_{sz}^{(r)}}
{\displaystyle P_{pz}^{(i)}},\qquad
T_{pp}=\frac{\displaystyle P_{pz}^{(t)}}
{\displaystyle P_{pz}^{(i)}},\qquad
T_{sp}=\frac{\displaystyle P_{sz}^{(t)}}
{\displaystyle P_{pz}^{(i)}}.
\end{array}
\end{equation}
The meanings of subscripts and superscripts in (3) are the same
as in (2).
The quantities $r_{sp}$, $t_{sp}$, $R_{sp}$, and $T_{sp}$ are
obviously the polarisation conversion coefficients.

Since the procedure of the solution of the electromagnetic problem is
quite standard, although rather cumbersome, we do not present here
explicit overlenghty expressions for magneto-optical and power
conversion coefficients.

It is well known [6] that the magnitudes of wave vector $k_x$ of the CPs
in the plane of $2D$ system lie in the  range $0<k_x<\omega\sqrt{\max(\varepsilon_1,\varepsilon_2)}/c$.
An external EW incident at an angle $\theta$ excites in the $2D$ system force
oscilations with longitudinal wave vector
$k_x=\omega\sqrt{\varepsilon_1}\sin{\theta}/c$.
In order to be able to investigate the entire range of variation
of the CP wave vector values, we assume $\varepsilon_1>\varepsilon_2$.
It is obvious that for $\theta>\theta_R$, where
$\theta_R=\sin^{-1}({\varepsilon_2/\varepsilon_1})^{1/2}$,
the regime of total internal reflection of the EW from
the media interface takes place.

The most interesting is the behaviour of polarisation conversion
coefficient $R_{sp}$ at the CP resonance. The corresponding curves for
the case where there is no electron scattering in the $2D$ system
($1/\tau=0$) are presented in Fig.2. The remaining parameters used in the
calculations are characteristic for GaAs/AlGaAs heterostructures with a
$2D$ electron gas. The choice of values of the dc magnetic field that
correspond to the results presented is dictated by the condition of
resonant excitation of the CPs $\omega\simeq\omega_c$
[6]. It follows from Fig.2 that $R_{sp}$ increases substantially in the
total internal reflection region and reaches
unity at a certain angle of incidence $\theta>\theta_R$.
For comparison, we
indicate that far from resonant values of the dc magnetic field the
calculations give a polarisation conversion coefficient
$R_{sp}$ less than $10^{-
3}$ for any angle of incidence.
At resonance (Fig.2) the extremely high polarisation conversion
coefficient remains in a quite wide range of variation of the
angle of incidence of the EW.
This is explained by the
comparatively weak dependence of the CP
frequency on the in-plane wave vector
$k_x=(\omega\sqrt{\varepsilon_1}/c)\sin{\theta}$ [6]. As it is evident
from Fig.2, polarisation conversion is quenched for $\theta=\theta_R$
and $\theta=\pi/2$. The physical explanation for this fact is that in the
both cases the total electric field at the $2D$ electron layer is
perpendicular to the $2D$ system plane [8]. Therefore, the external EW does
not interact with in-plane excitations of $2D$ system at these angles.

In Fig.3 the dark tone indicates $\omega-\theta$ regions in which the
polarisation conversion coefficient $R_{sp}>0.99$ for various values of
the surface electron density in $2D$ system. The figure also shows the CP
dispersion curves $\omega[k_x(\theta)]$ for
the same values of the surface density. For a low electron density the
total polarisation conversion occurs near the cyclotron-polariton frequency
($\omega\simeq\omega_c$) for any angle of incidence of the EW.
This corresponds to the case of weak coupling when the external
wave induces comparatively low electric currents in the $2D$ system. As the
electron density increases, the coupling of the external EW to the $2D$
system grows, and the frequency range in which almost total polarisation
conversion occurs broadens. Under strong coupling the total polarisation
conversion occurs at resonant frequencies different from the CP
eigen-frequencies.

The state of polarisation of reflected wave may be determined
from the ratio of the magneto-optical complex reflection
coefficients (2)
$r_{sp}/r_{pp}=(|r_{sp}|/|r_{pp}|)\exp (i\delta)$.
The magneto-optical rotation angle $\psi$
and ellipticity $\varepsilon$ of reflected
wave are calculated
by the formulas [9]:
\begin{equation}
\begin{array}{rcl}
{\rm tg}2\psi &=&{\rm tg}2\varphi\cos\delta,\\
\sin 2\varepsilon&=&\sin 2\varphi\sin\delta,
\end{array}
\end{equation}
where $\varphi={\rm tg}^{-1}(|r_{sp}|/|r_{pp}|)$.
In Fig.4 the rotation angle and ellipticity
are shown in the frequency range across the CP magneto-optical
resonance in the total reflection regime.
In the absence of the electron scattering in $2D$ system
the ellipticity becomes zero at
the resonance and the rotation angle reaches the values of $\pm\pi/2$.
These correspond to the total polarisation conversion conditions. The
rotation angle experiences a jump of $\pi$ and the handedness of
elliptical polarisation reverses its sign at the resonance. The electron
scattering in a real $2D$ electron system leads to a gradual but still very
steep transition of the rotation angle from positive to negative values
across the resonance (see Fig.4).

The effect of electron scattering in $2D$ system on the CP assisted
polarisation conversion is demonstrated in Fig.5. It is evident that
if the coupling between external EW and the CPs in
the $2D$ system is weak (low electron density), the electron scattering
suppresses the polarisation conversion effect almost completely like
in the case of MPP assisted polarisation conversion in ATR geometry.
At the same time, a high polarisation
conversion efficiency at CP resonance
remains even in the presence of the electron
scattering if the electron density in the $2D$ system is high.
Because of that the CP assisted
polarisation conversion at high electron densities in $2D$ system is much
more profitable from the practical point of view as compared to
MPP assisted one.

Note that the solution of the problem of the incidence
of an $s$-polarized wave onto a $2D$ electron system gives the same
values for the polarisation conversion coefficients
($R_{ps}=R_{sp}$) in the total reflection regime. This attests to a
reciprocal character of the polarisation conversion process for
$\theta\geq\theta_R$.

The thank Yu.A.Kosevich for calling our attention to the
problem under consideration.
This work was supported by the Federal Target Program "Integration"
(Project 696.3) and by the Russian Foundation for Basic Research
(Grant No. 00-02-16440).

\begin{center}
{\bf References}
\end{center}

\begin{enumerate}
\item
R.F.O'Connell, G.Wallace, Phys. Rev. B., 26 (1982) 2231.
\item
V.A.Volkov and S.A.Mikhailov, JETP Lett., 41 (1985) 476.
\item
V.A.Volkov, D.V.Galchenkov, L.A.Galchenkov et al., JETP Lett.,
43 (1986) 326.
\item
L.A.Galchenkov, I.M.Grodnenskii, M.V.Kostovetskii et al., Fiz.
Tekh. Poluprovodn., 22 (1988) 1196 [Sov. Phys. Semicond.,
22 (1988) 757].
\item
M.I.Bakunov and S.I.Zhukov, Pis`ma Zh. Tekh. Fiz.,
16, 1 (1990) 69 [Sov. Tech. Phys. Lett., 16 (1990) 30].
\item
V.V.Popov, T.V.Teperik, and G.M.Tsymbalov,
JETP Lett., 68 (1998) 210.
\item
Yu.A.Kosevich, Solid State Commun., 104 (1997) 321.
\item
V.V.Popov and T.V.Teperik, JETP Lett., 70 (1999) 254.
\item
M.Born and E.Wolf, {\sl Principles of Optics} (6th edn).
Oxford: Pergamon Press, 1980.
\end{enumerate}

\begin{center}
{\bf Figure captions}.
\end{center}

Fig.1. The structure under consideration
and coordinate system. $\varepsilon_1$ and
$\varepsilon_2$ are dielectric constants of the
media.

Fig.2. Coefficient of conversion of the polarisation
in the reflected wave vs the angle of incidence:
a) $N_s=2\times 10^{12}$ cm$^{-2}$;
$B_0$ (kG): 44 ({\sl 1}),
45 ({\sl 2}), 46 ({\sl 3}), 46.5 ({\sl 4}), 47 ({\sl 5});
b) $B_0=46$ kG;
$N_s(10^{12}$ cm$^{-2})$: 0.8 ({\sl 1}),
0.9 ({\sl 2}), 1 ({\sl 3}), 2 ({\sl 4}), 3 ({\sl 5}).
$\omega/2\pi c=60$ cm$^{-1}$,
$\varepsilon_1=12.8$, $\varepsilon_2=1$.

Fig.3. Regions of the parameters $\omega-\theta$ with the
polarisation conversion efficiency $R_{sp}>0.99$
and dispersion curves for cyclotron-polaritons for
$\varepsilon_1=12.8$, $\varepsilon_2=1$,
$B_0=46$ kG, and  $N_s(10^{12}$ cm$^{-2})$: 0.2 ({\sl 1}),
1 ({\sl 2}), 2 ({\sl 3}), 3 ({\sl 4}).

Fig.4. Rotation angle (solid curves) and ellipticity (dashed curves)
of the reflected wave vs frequency
across the CP resonance with no electron scattering (curves {\sl 1})
and with electron scattering (curves {\sl 2}, $\tau=10^{-12}$ s)
in $2D$ system.

Fig.5. Curves {\sl 1} and {\sl 3} correspond to the case when there
is no electron scattering in the $2D$ system
($1/\tau=0$);
curves {\sl 2} and {\sl 4} correspond to
$1/\tau=10^{11}$ s$^{-1}$. Curves {\sl 1} and {\sl 2}:
$N_s=2\times 10^{12}$ cm$^{-2}$,
$B_0=46$ kG;
curves {\sl 3} and {\sl 4}:
$N_s=2\times 10^{11}$ cm$^{-2}$,
$B_0=45.3$ kG. All other parameters are the same as in Fig.2.

\end{document}